\documentclass[aps,nofootinbib]{revtex4} 

  \usepackage{relsize} 
 \usepackage{hyperref}
  \usepackage{amsmath,amsfonts,amssymb}
  \hypersetup{colorlinks=true,urlcolor=magenta,filecolor=magenta,linktoc=page,citecolor=blue,linkcolor=blue}
  \usepackage{graphicx,epstopdf}
  \usepackage[inline]{enumitem}

\begin{document}
 	
\title{Reconstructing a non-linear interaction in the dark sector with cosmological observations}
 	
 \author{Jiangang Kang}
 \email{kjg@nao.cas.cn}
 \affiliation{National Astronomical Observatories, Chinese Academy of Sciences, Beijing 100101, China}
 \affiliation{CAS Key Laboratory of FAST,National Astronomical Observatories, Chinese Academy of Sciences, Beijing 100101, China}
 \affiliation{School of Astronomy and Space Science,University of Chinese Academy of Sciences, Beijing 100049, China}

\begin{abstract}
 		In this work we model two non-linear directly interacting scenarios in dark sector of  the universe with the dimensionless parameter $\alpha$ and $\beta$, which dominate the energy exchange  between  dark energy and dark matter. The central goal of this investigation is to research the interacting model and discuss the cosmological implications based on the current observational datasets. The class of the interaction is generally characterized by a coupling function $Q\propto H(z)\rho_x$, $x$ denotes the energy density of dark matter or dark energy. The constrained results we obtained indicate that the direct interaction in cosmic dark sector  is  favored  by  various observational data and the key effects on CMB power spectrum and linear matter power spectrum appear compared to $\Lambda$CDM standard paradigm. Finally, we discuss in depth the effect of  different neutrino mass hierarchy on matter power spectrum and the variation of the ratio of CMB temperature  power spectrum $C_{\ell}^{TT}$ and matter power spectrum $P(k)$ when the $\Delta N_{eff}= N-3.046$ from 0.5 to 2, respectively.
\end{abstract}
\maketitle
$~~~~~~$\textbf{Keywords}: dark energy; dark matter; cosmological observations; cosmological parameters; neutrinos.

\section{Introduction}\label{intro}

 The accelerated expansion of the late-time Universe since  was discoveried by observations of type Ia supernovae \citep{1998AJ....116.1009R,1999ApJ...517..565P}, understanding the  physical nature behind the
acceleration is one of the fundamental topics in modern cosmology. 

An exotic species of the energy with the equation of state $w_d^{eff} =P_d/\rho_d < -\frac{1}{3}$ dubbed as  dark energy (DE) has been introduced \citep{2003RvMP...75..559,2006IJMPD..15.1753C}  to explain the expansion of the Universe. Based on Einstein's theory of General Relativity(GR), the dark energy is represented by the cosmological constant $\Lambda$, which combinates with the pressureless cold dark matter (CDM) have regarded as a standard model for modern cosmology.

Many predictions from the standard $\Lambda$ cold dark matter($\Lambda$CDM) theory have been successfully verified by a variety of astronomical observations. However, there are still several challenging  puzzles which need to be addressed, (i)
the so-call fine tuning problem, i.e. the result of the vacuum energy density based on  quantum theory reaches some 120 orders of magnitude larger than that inferred from the cosmological constant $\Lambda$ \citep{2011CoTPh..56..525L,1995A&A...301..321W}; (ii) coincidence puzzle: is that the energy density of DM and DE  are approximate one order of magnitude, although the two quantities do not share similar evolution laws and have different rates of evolution as the universe expands, so why do they happen to be of the same order of magnitude right now? \citep{2006IJMPD..15.1753C,2005PhLB..624..141W,2002PhRvD..65f3508T, 2003PhRvD..67h3513C, 2016cosp...41E1661R, 1997JOSAB..14.1275P,1999PhRvL..82..896Z}; (iii) the Hubble constant tension: the$\ H_0=67.4\pm 0.5 kms^{-1}Mpc^{-1}$  derived from Planck 2018( P18) Cosmic Microwave Background (CMB)  \citep{collaboration2018planck,2017NatAs...1..569D} determination in $\Lambda$CDM  is less than 4.4$\sigma$ error that$\ H_0=74.03 \pm 1.82 kms^{-1}Mpc^{-1}$ obtained directly from the local distance ladder measurements of SN Ia (R19) \citep{2019ApJ...876...85R};
(iv) the estimated root-mean-square mass fluctuation amplitude $\sigma_8$ in 8$h^{-1}Mpc$ spherical volume and matter energy density $\Omega_m$ from the Planck 2018 CMB measurements \citep{collaboration2018planck,2013MNRAS.430.2200K} under the $\Lambda$CDM cosmology is different from that derived from weak lensing effects based on the KiDS survey \citep{2017MNRAS.471.4412K,2017MNRAS.465.1454H,2018MNRAS.474.4894J}. Therefore by measuring the growth rate of structure in the distribution of galaxies as function of redshift, one can place constraints on gravity, and test if dark energy could be due to deviations from GR \citep{2008Natur.451..541G}.

In order to address these discrepancies, a reasonable solution is recently introduced  as a novel mechanism beyond the canonical $\Lambda$CDM model. This modification to dark energy, which is treated as a smooth non-clustering perfect fluid, may directly interact with dark matter without gravity action. Such an idea has been extensively investigated by a number of phenomenological models, which show that the interaction between DE and DM can alleviate the coincidence
problem, and the model predictions have been passed various observational tests. In particular, many parameterized models assume dynamical couplings between DE and DM, and show that the current observational  measurements support a nonzero interplay behavior for the cosmic dark sector\citep{2014PhRvL.113r1301S, 2014PhRvD..90h3532Y, 2014PhRvD..89h3517Y, 2016PhRvD..94b3508N, 2017PhRvD..95d3513V,2016JCAP...10..007Y,2019PhRvD.100b3515V,2019MNRAS.482.1007Y,2019MNRAS.488.3423M,10.1093/mnras/staa213,2005PhRvD..72l3523B,2018ChPhC..42i5103G, van_de_Bruck_2018}. Generally, these dynamical coupling scenarios are described with a functional coupling term $Q$. In this paper, we propose two novel coupling scenarios of DE and DM, and then put constraints on the relevant parameters within the coupling model $Q_1$ and $Q_2$ with the help of mainstream cosmological the combination of datasets and  finally  we research the  impact of two interaction models on the   $C_\ell^{TT}$ and matter power $P(k)$ spectra on large scale of universe. Beyond that, we comment on different the neutrino mass hierarchy problem \citep{2018PhRvD..98h3501V,2013arXiv1307.4738B,2001hep.ex....9033K}, i.e. either $m_1<m_2\ll m_3$, normal hierarchy(NH) or $m_3\ll m_1< m_2$, inverse hierarchy(IH) and  analyse the ratios with $\Lambda$CDM when setting $\Sigma m_{\nu}=0.06eV$ and  comment on some crucial hints about the neutrino mass splittings can be demonstrated  when investigating these IDE models with the current cosmological observations. From the viewpoint of phenomenology, the given assumptions made for $Q$  function should be tested with a number of cosmological probes \citep{2020SCPMA..6320401F}.

This paper is structured as follows: The scenarios of the interaction IDE1 and IDE2 in dark sector we propose are introduced and from which we derive the relevant evolution equations in Section \ref{mod}. In Section \ref{data}, we present the SN Ia, BAO, CMB and H(z) data
combinations adopted in this work as well as the analysis method in each fitting process with these datasets. The eventual results about the parameters and the $C_{\ell}$ and $P(k)$ spectra are obtained from the  IDE1 and IDE2 models in Section \ref{res}. Finally, we draw our conclusions on two interacting schemes  in Section \ref{cons}. We use natural units throughout $c=8\pi G =1$.
 	
\section{The model of the interaction in dark sector}\label{mod}
Considering a homogeneous and isotropic Universe with a spatially flat Friedmann-Lematre-Robertson-Walker(FLRW) spacetime 
whose line element follows the form:
\begin{equation}\label{ds2}
 ds^2 = -dt^2 + a^{2}(t) \left(dr^2 + r^2 d\theta^2 + r^2sin^{2}\theta d \phi^2\right),
\end{equation}
 where $a(t)$ is cosmic scale factor, $t$ is cosmic time and ($r,\theta,\phi$) is comoving spherical coordinate system. For a uncoupled standard cosmological model (SM) usually contains radiation, baryons, dark matter and  dark energy, their  perfect  fluid  energy–momentum tensor satisfies $ T_{\mu\nu} = pg_{\mu\nu}+(\rho+p)U_{\mu} U_{\nu} $ respectively, 
 	where $U_{\mu,\nu}$ stands for four velocity vector. When introducing the directly interaction of dark constituents, one can understand the  conservation equation of total energy momentum tensor as $\bigtriangledown_{\mu} (T^{\mu\nu}_c+T^{\mu\nu}_d)=0$ based on the Einstein's field equation $G_{\mu\nu} = T_{\mu\nu}^c+T_{\mu\nu}^d$, $c,d$ denotes dark matter and dark energy while the radiation and baryons independently evolve as own mode, respectively. Thus, the FLRW evolution equation for this framework can be written as
 	\begin{align}  
 	&3H^2 =\rho_d+\rho_c,\label{energy1}\\
 	&2\dot{H} + H^2 =-p_d,\label{energy2}
 	\end{align}
 	where an overdot refers to a derivative w.r.t the cosmic time $t$, $H =\dot{a}/a$ is the Hubble parameter, $\rho_c$, $\rho_{d}$ are the matter density of dark matter  and dark energy  and $\ p_c$,$\ p_{d}$ represents the corresponding pressure of dark matter and dark energy, respectively. Dark matter is usually treated as a pressureless dust matter,$\ p_c=0$ while dark energy is taken  as a negative pressure fluid $\ p_d =w_d\rho_d$ with the effective equation of state(EoS) $w_d$. For instance, if the DE is the smooth vacuum or $\Lambda$, i.e. $w_d$ = -1. The physical behavior of the effective dark fluid when the presence of interaction in dark sector will be different in light of the sign of the coupling function $Q$. Thus, the whole energy conservation equation for the interaction  can be read as the following forms:
 	\begin{align}
 	\dot{\rho}_c + 3H(1+w_c^{eff})\rho_c &= Q \, \label{conv1} \\
 	\dot{\rho}_{d} + 3H(1+ w_d^{eff})\rho_{d} &= -Q \, \label{conv2}
 	\end{align}
 	For a interacting dark fluid of Universe, the sign and strength of the coupling function $Q\neq 0 $ determines the effective rate of energy conversion between DM and
 	DE. In the case of $Q<0$, meaning that the energy transfer from dust DM to DE, while for $Q> 0$, oppositely  which implys the DE likewise decay into pressureless DM. As a phenomenological model, it is usually  assumed that the interaction term $Q$  is proportional to the product of energy density of a dark fluid and the Hubble parameter, $Q\propto \rho_x H$. Conventionally, the coupling strength $Q$ can be tested by a number of observational results with a linear dependence on the energy density of dark elements  as presented in the mathematical formula \citep{2009JCAP...07..034G,2008JCAP...05..007Q,2008PhRvD..78b3505B,2009AIPC.1122..197B,2007CQGra..24.5461Z,Bachega_2020}:  
 	\begin{equation}
 	Q = 3H(\alpha_c\rho_c +\alpha_d\rho_{d}),\label{int1}
 	\end{equation}
 	where $\alpha_c$ and $\alpha_d$  are dimensionless constants dominating the strength of the DM-DE interaction. In absence of the interaction,namely $Q=0$, the model restore to $\Lambda$CDM cosmology. The redshifted energy density  evolution equations for any coupling term $Q$ and the effective EoS $w^{eff}$ for dark sector can be  deduced as:
 	\begin{align}
 	\ &\rho_c =  \rho_{c0}\left[\int_{a_0}^{a}(\frac{Q}{aH})da\right] ,\label{density1}\\
 	\rho_d = & \rho_{d0}\left[a^{-3(1+w_d)} - \int_{a_0}^{a}(\frac{Q}{aH})\right]da,\label{density2}\\
 	\ w_c^{eff}& = \frac{Q}{3H\rho_c}, \quad w_d^{eff} = w_d - \frac{Q}{3H\rho_d},\label{eos}
 	\end{align}
 	where $w_c^{eff}$ and $w_{d}^{eff}$ are respectively termed as the effective equation of state for CDM and DE, $\rho_{c0}$ and $\rho_{d0}$ denote cold dark matter and dark energy current matter density, and $a_0$ is today cosmic factor, $Q$ stands for any coupling function of dark sector,respectively,
 	
 	The origin of instabilities depend on different signs of $\alpha_c$ and $\alpha_d$ and effective EoS $w^{eff}$ under the interacting dark energy (IDE) scenarios\citep{2014PhRvD..90f3005L,2014PhRvD..90l3007L,2017JCAP...05..040G,2017SCPMA..60e0431Z,2019ApJ...876..125D,2018JCAP...09..019Y}. In this work, we propose originally two interacting scenarios as following physical mechanism: in which the coupling function $Q(t)$ can be defined as:
 	\begin{align}
 	IDE1: Q_1 = 3H\alpha\rho_d\left[1+\beta ln(1-\frac{\rho_c}{\rho_d})\right],\label{eq:IDE1}\\
 	IDE2 : Q_2= 3H\alpha\rho_d\left[1+\beta sin(\frac{\rho_c}{\rho_d})\right],\label{eq:IDE2}
 	\end{align}
 	where $\alpha $ and $\beta $ are dimensionless coupling parameter governing the interaction strength of the dark sector and then one can expands $Q$ function  around the $\rho_c/\rho_d = 1$ in the manner of Taylor series and the first order approximation we take for the interaction model IDE1 and IDE2 and thus $Q_1$ and $Q_2$ can be rewritten as a non-linear expression:
 	\begin{equation}
 	Q_1\approx 3H\alpha\rho_d(1-\frac{\beta\rho_c}{\rho_d})= 3H\alpha(\rho_d-\beta\rho_c),\label{eq:IDE1e}\\ 
 	\end{equation}
 	and 
 	\begin{equation}
 	Q_2\approx 3H\alpha\rho_d(1+\frac{\beta\rho_c}{\rho_d})=3H\alpha(\rho_d+\beta\rho_c),\label{eq:IDE2e}
 	\end{equation}
 	Then we further derive the effective EoS Eq.\ref{eos} in term of above interaction function via $r=\rho_c/\rho_d$ for IDE1 model:
 	\begin{equation}
 	\ w_c^{eff}= \alpha(\frac{1}{r}-\beta), \quad w_{d}^{eff} = w_d -\alpha(1-\beta r),\label{eos1}
 	\end{equation}
 	meanwhile for IDE2:
 	\begin{equation}
 	\ w_c^{eff}= \alpha(\frac{1}{r}+\beta), \quad w_{d}^{eff} = w_d -\alpha(1+\beta r),\label{eos2}
 	\end{equation}
 	\begin{figure}
 		\begin{center}
 			\includegraphics[width=0.49\textwidth]{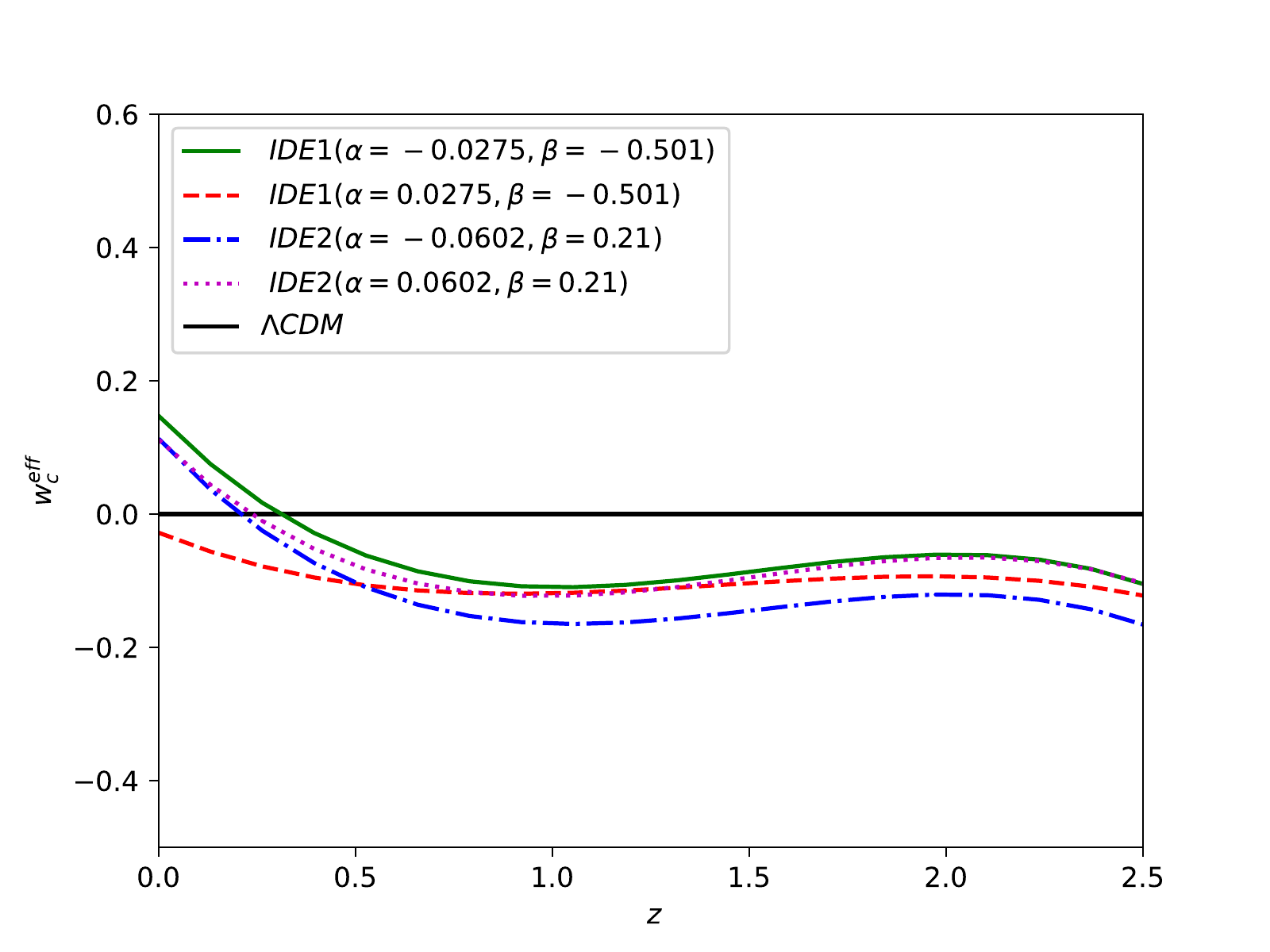}
 			\includegraphics[width=0.49\textwidth]{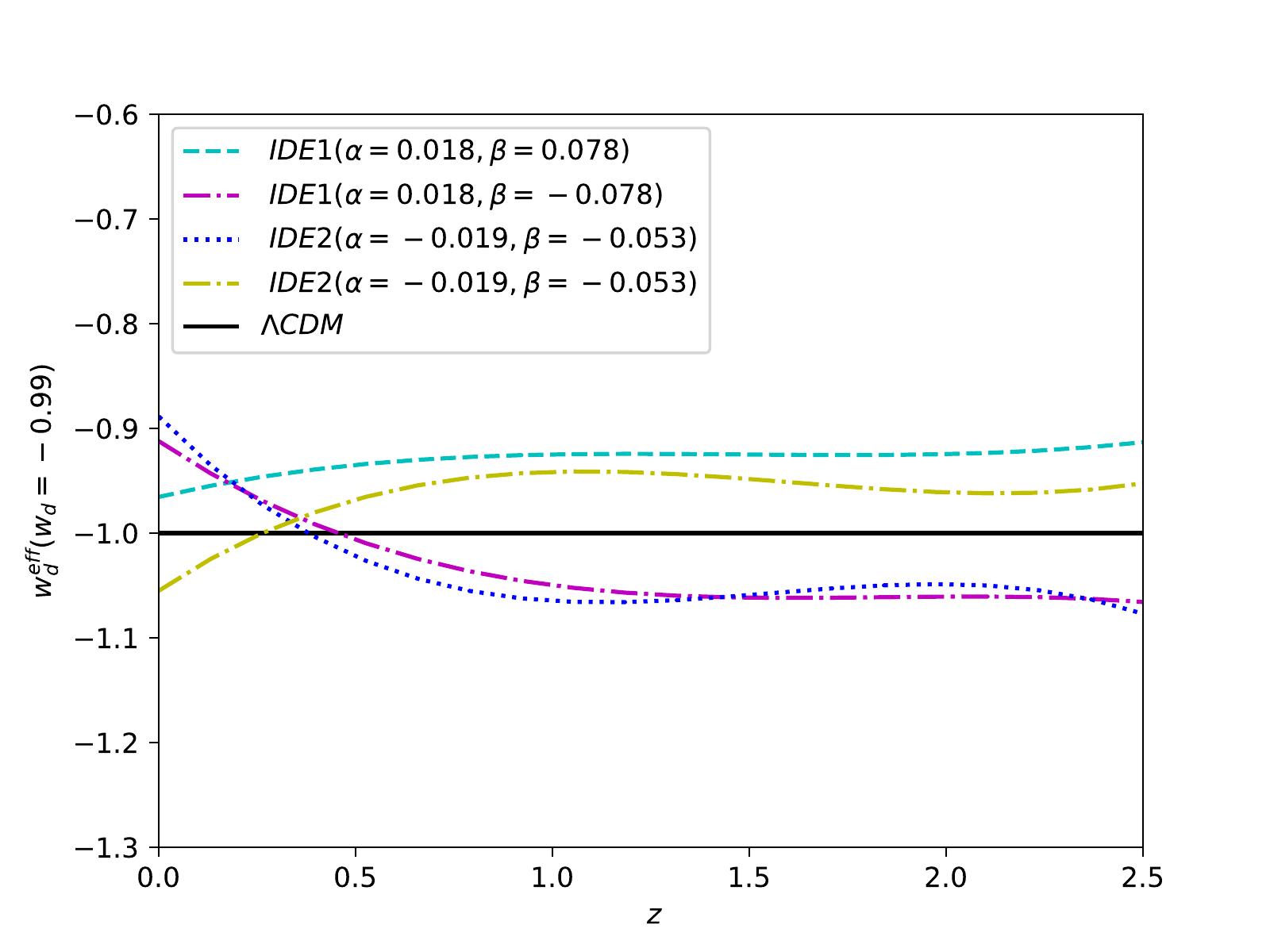}
 			\caption{\label{fig:eos1}The figure shows the redshift  evolution of the effective equations of state for DE(top) and CDM (below) under the interaction models $Q_1$ and $Q_2$ when taking different two groups value for  $\alpha$ and $\beta$,respectively.}
 		\end{center}
 	\end{figure}
 	Fig.\ref{fig:eos1} shows the evolution behavior of $w_c^{eff}$
 	and $w_d^{eff}$ with respect to redshift $z$ for the two interaction schemes $Q_1$ and $Q_2$,respectively. The left panel of Fig.\ref{fig:eos1} displays the evolution of $w_c^{eff}$ as well as the evolution of $w_d^{eff}$ in the right panel  taking same values of the $\alpha$ and $\beta$ for IDE1 as well as IDE2  when fixing $w_d=-0.99$, one can find the effective equations of state for dust DM and DE are sensitive to the interaction parameter $\alpha$ and $\beta$ and the evolution of $w_c^{eff}$ and $w_d^{eff}$ for IDE1 and IDE2  maintain stable  for $z > 1.5$ around the $w_d =-1$ from the picture. As for IDE1 we see, the energy transition occurs from DM to DE $Q < 0$ ($\alpha<0$ and $\beta >0$) while $Q > 0$ ($\alpha>0$ and $\beta < 0$) energy flows from DM to DE after $z>0.5$, while for IDE2 case, the energy flows between DE and DM  opposite direction compared to IDE1 case in term of the various sign of $\alpha$ and $\beta$.
 	
 	Now, the main perturbative equations of energy density $\delta$ and velocity $v$ under the any interaction  $Q$ of DE and DM model in synchronous gauge for dark energy become \citep{1995ApJ...455....7M,2009PhRvD..79d3526J,2010JCAP...11..044G,2009JCAP...07..034G,2013PhRvD..88b3531S,2019PDU....23..261F,2010JCAP...10..014B,2008IJMPD..17.1229X, 2016JCAP...12..009M}:
 	\begin{eqnarray}
 	\delta^{\prime}_{d} =\dfrac{a}{\bar{\rho}_{d}}{Q}_{d} -\dfrac{a{Q}}{\bar{\rho}_{d}}\left[\delta_{d}
 	+3\mathcal{H}\left(c^{2}_{(s)d}-w_d\right)\delta v\right],
 	\label{dep}
 	\end{eqnarray}
 	\begin{eqnarray}
 	v^{\prime}_{d}=\dfrac{a{Q}}{\bar{\rho}_{d}}\left[v_d-\left(1+c^{2}_{(s)d}\right) v_d\right]-\dfrac{a}
 	{{\rho}_{d}} \left[\dfrac{c^{2}_{(s)d}\rho_{d}}{a}\delta_{d}-{Q}{v}\right].
 	\label{devp}
 	\end{eqnarray}
 	
 	Similarly, as for the cold dark matter perturbations, $w_c =0$ leading to $c^{2}_{(a)c}={c}^{2}_{(s)c}=0$, its perturbative formula of energy $\delta_c^{\prime}$ and speed $v^{\prime}_c$ can be read respectively as:
 	\begin{eqnarray}\label{cdmvp}
 	\delta^{\prime}_{c}=k^{2}v_{c}-\dfrac{h^{\prime}}{2}-\dfrac{a}{\bar{\rho}_{c}}{Q}+\dfrac{a{Q}}{\bar{\rho}_{c}}\delta_{c}.\\
 	v^{\prime}_{c}=-\mathcal{H}v_{c}-\dfrac{a{Q}}{\bar{\rho}_{c}}\left(v-v_{c}\right).
 	\end{eqnarray}
 	here $\delta_i= \delta \rho/\rho$ ($i = c,d$) and $\mathcal{H}$ = $aH$ is comoving Hubble parameter. 
 	
\begin{figure}
 		\centering
 		\includegraphics[width=0.6\textwidth]{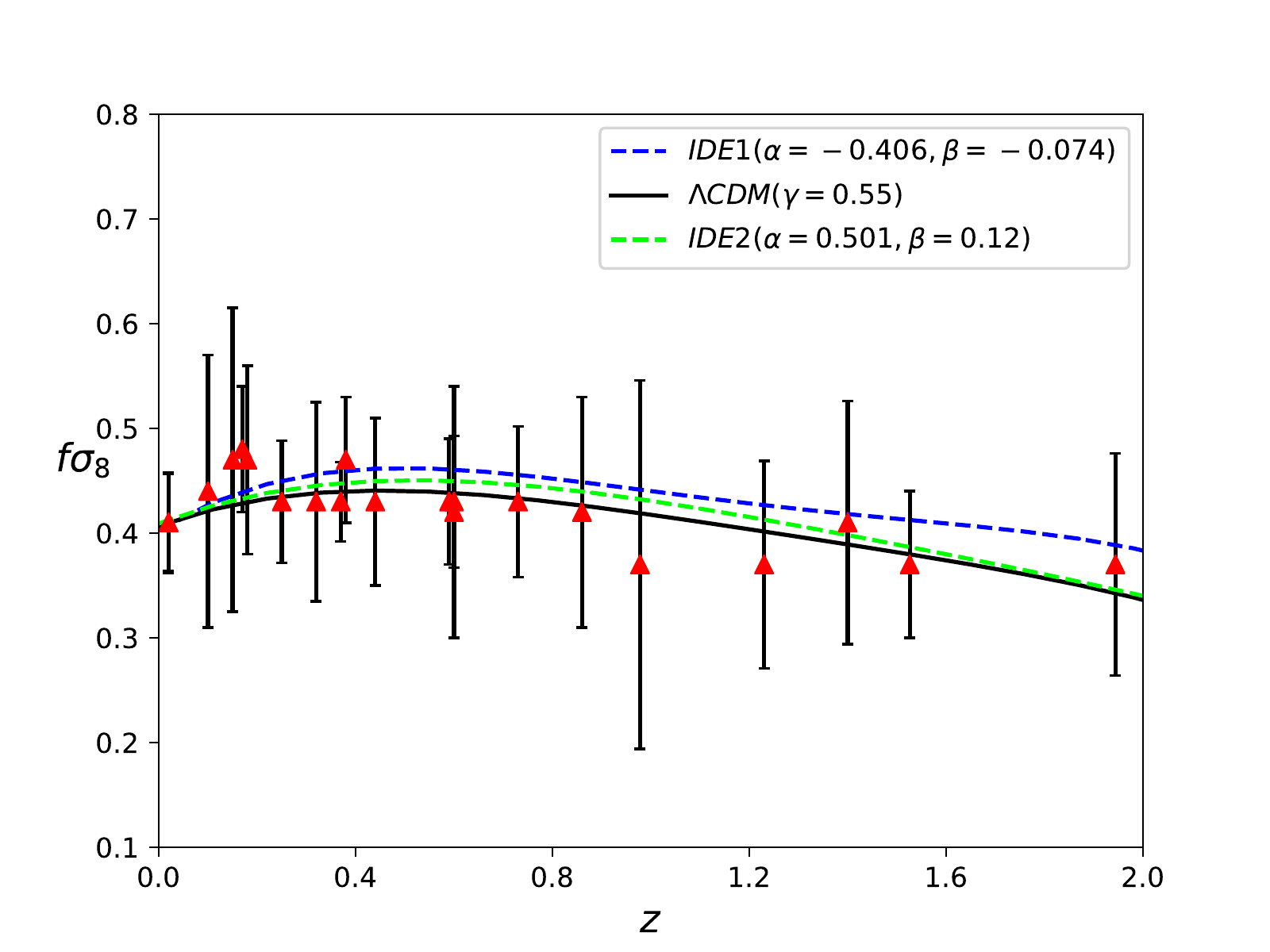}
 		\caption{\label{fig:growth}The linear evolution of  $f\sigma_8$ as the function of redshift $z$ for interaction models $Q_1$ and $Q_2$ (dash blue and lime line,black solid line stands for $\Lambda$CDM model, respectively). Errorbar datapoints are employed in Table I of \citep{2018PhRvD..98h3543S}}
 		
 	\end{figure}

 	Besides, we discuss the contribution of matter overdensity $\delta$  that resulted from the interaction  between dark energy and dark matter. In term of the general definition of the growth rate of matter perturbations: $f_c= \frac{dln\delta_c}{dlna}$=$\delta'$/$\delta_{c}$  and for an arbitrary interaction function $Q$ \citep{2020PhRvD.101j3507J,2020PDU....3000616B,10.1093/mnras/staa213}, the growth rate  can be understood as:
 	\begin{eqnarray}\label{growth-rate}
 	\delta''_c+\left(1-\frac{Q}{H\rho_c}\right)\mathcal{H}\delta'_c 	=\frac{3}{2}\mathcal{H}^2\Omega_b\delta_b+\frac{3}{2}\mathcal{H}^2\Omega_c\delta_c \Bigg\{1 +\frac{2}{3}\frac{\rho_d}{\rho_c}\frac{Q}{H\rho_c} \nonumber \\
 	\Bigg[ \frac{\mathcal{H}'}{\mathcal{H}^2},
 	+1-3w_d+\frac{w_d}{\mathcal{H}(1+w_d)}	+\frac{Q}{H\rho_c}\left(1+\frac{\rho_d}{\rho_c}\right) \Bigg] \Bigg\}, 
 	\end{eqnarray}
 	It's difficult to solve the analysis expression of $f_c$ in Eq.\ref{growth-rate} but we can fit it using the format $f(z)= \Omega_{\rm m}^{\gamma}$ ($\gamma \approx 0.55$ in $\Lambda$CDM cosmology\citep{2007APh....28..481L,Hudson_2012}) with observations combinations as shown in Fig.\ref{fig:growth} and from the plot, we find that growth formation of the CDM is much sensitive to various interaction model $Q$ and coupling factors especially after redshfit $z=2$  when DE begins to dominate cosmic energy ingredient and drives the accelerated expansion of the universe . 
 	
 	In this work, we investigate cosmological dynamics with the flat spacetime due to the above Eqs.\ref{ds2}--\ref{eq:IDE2e}, neglecting the radiation density $\Omega_{r0}\approx 10^{-5}$, therefore assuming the matter density relation $\Omega_m$ +$\Omega_d$ = 1 for IDE1 case can be rewritten:
 	\begin{eqnarray}\label{id1}
 	H(z)^2/H_0^2=\Omega_{d}\left(\frac{w}{\alpha+\beta}(1+z)^{3}+\frac{1}{\alpha+\beta}(1+z)^{3(1+w+\alpha)}\right) +\Omega_{m}(1+z)^{3},
 	\end{eqnarray}
 	For the IDE2 scheme: 
 	\begin{eqnarray}\label{id2}
 	H(z)^2/H_0^2=\Omega_{m}\left(\frac{w}{\alpha +\beta}(1+z)^{3(1+w)}+\frac{1}{\alpha+\beta}(1+z)^{3(1-w-\alpha)}\right)\
 	+\Omega_{d}(1+z)^{3(1+w)}
 	\end{eqnarray}
 	where $\Omega_{m}$ and $\Omega_{d}$ are current energy density of dark matter and dark energy.

\section{Data and Methodology}\label{data}
In this section we introduce the mainstreams cosmological probes and the likelihood function $\mathcal{L}\propto e^{-\chi^2/2}$ used in this work to extract the posterior distribution functions (PDFs) of each parameter within  IDE1 and IDE2 framework.
\subsection{SN Ia data}
The Type Ia Supernovae (SN Ia) play a key role in unveiling the character of the dark energy  as the "standard candles". We utilize the Pantheon compilation consisting of 1048 SN Ia samples covering redshift range $0.01<z<2.3$ \citep{2018ApJ...859..101S}. The $\chi^2$ distribution for the SN Ia can be written as
\begin{equation}
\chi^2_{\rm SN}=\Delta {\bf u}^T\cdot{\bf C^{-1}}\cdot\Delta {\bf u}.
\end{equation}
Here $\Delta{\bf u}={\bf u}_{\rm obs}-{\bf u}_{\rm th}$, ${\bf u}_{\rm obs}$ and ${\bf u}_{\rm th}$ are the vectors of observational and theoretical distance modulus, respectively,and ${\bf C}$ is the covariance matrix.$\ u_{\rm th}=m_{\rm th}-M$ , where $ u_{\rm th}$ is the theoretical apparent magnitude, it can be obtained as
\begin{eqnarray}
\ &u_{th}=5\rm {log}_{10}D_{\rm L}(z)+25, \label{eq:mu_M}
\end{eqnarray}
where $M$ is the absolute magnitude of SN Ia as a nuisance parameter, $D_{\rm L}(z)=(1+z)D_{\rm C}(z)$ is the luminosity distance in unit of $\ Mpc$. The covariance matrix $\bf C$ takes the form as
\begin{equation}
{\bf C} = {\bf C}_{\rm stat} + {\bf C}_{\rm sys},
\end{equation}
where ${\bf C}_{\rm stat}$ is the vector of statistic error, ${\bf C}_{\rm sys}$ is the systematic covariance matrix for the data \citep{2018ApJ...859..101S}.
\subsection{BAO data}
\begin{table*}
	\vspace{1mm} 
	\begin{center}
		\caption{The 12 BAO data used in this work redshift from $z=0.1$ to 2.4.}\label{tab:bao} 
		\begin{tabular}{c|c|c|c|c|c}
			\hline  \hline
			Redshift &Measurement& Value &$r_{\rm s, fid}$&Survey & Reference\\[0.8ex]
			\hline
			0.106 &$r_{\rm s}/D_{\rm V}$& 0.3360$\pm$0.015 &-- &6dFGS  &\citep{2011MNRAS.416.3017B}\\
			0.15 &$r_{\rm s}/D_{\rm V}$&0.2239$\pm$0.0084  &-- &SDSS DR7 &\citep{2015MNRAS.449..835R}\\
			0.32 &$r_{\rm s}/D_{\rm V}$&0.1181$\pm$0.0024  &-- &BOSS LOW-Z &\citep{2014MNRAS.441...24A}\\
			0.57 &$r_{\rm s}/D_{\rm V}$&0.0726$\pm$0.0007  &-- &BOSS CMASS &\citep{2012MNRAS.427.2132P}\\
			0.44 &$r_{\rm s}/D_{\rm V}$&0.0870$\pm$0.0042 &-- &WiggleZ &\citep{Blake2012}\\
			0.60 &$r_{\rm s}/D_{\rm V}$&0.0672$\pm$0.0031 &-- &WiggleZ &\citep{Blake2012}\\
			0.73 &$r_{\rm s}/D_{\rm V}$&0.0593$\pm$0.0020 &-- &WiggleZ &\citep{Blake2012}\\
			2.34 &$r_{\rm s}/D_{\rm V}$&0.0320$\pm$0.0013  &-- &SDSS-III DR11 &\citep{2015AA...574A..59D}\\
			2.36 &$r_{\rm s}/D_{\rm V}$&0.0329$\pm$0.0009  &-- &SDSS-III DR11 &\citep{2015JPhD...48R3001H}\\
			0.38 &$D_{\rm M}(r_{\rm s, fid}/r_{\rm s})$&1518$\pm$22  &147.78 &SDSS DR12 &\citep{2017MNRAS.470.2617A}\\
			0.51 &$D_{\rm M}(r_{\rm s, fid}/r_{\rm s})$&1977$\pm$27  &147.78 &SDSS DR12 &\citep{2017MNRAS.470.2617A}\\
			2.40 &$D_{\rm H}/r_{\rm s}$&8.94$\pm$0.22 &--        &SDSS DR12    &\citep{2017AA...608A.130D}\\
			\hline
		\end{tabular}
	\end{center}
	\vspace{-2mm}
\end{table*}

Table.\ref{tab:bao} lists the adopted quantities that derived from the different BAO measurements as the "standard ruler", $r_{\rm s}$ is the radius of the comoving sound horizon at the drag epoch as following:
\begin{equation}
r_{\rm s} = \int_0^{t_{\rm s}}c_{\rm s}\frac{{\rm d}t}{a},
\end{equation}
where $c_{\rm s}$ is the sound speed, $t_{\rm s}$ is the epoch of last scattering, $a$ is the scale factor. Since $r_{\rm s}$ are not sensitive to physics at low redshifts, we fix $r_{\rm s}=147.09\pm0.26$ measured from the Planck 2018 release \citep{collaboration2018planck}. The spherically averaged distance $D_{\rm V}$ is given by
\begin{equation}
D_{\rm V}(z) = \left[ D_{\rm M}^2(z)\frac{cz}{H(z)} \right]^{1/3},
\end{equation}
where $D_{\rm M}(z)=(1+z)D_{\rm A}$ is the comoving angular diameter distance, and $D_{\rm A}=D_{\rm C}/(1+z)$ is the physical angular diameter distance. $D_{\rm H}=c/H(z)$ is the Hubble distance.

The $\chi^2$ for the BAO data can be calculated by
\begin{equation}
\chi^2_{\rm BAO} =\Delta {\bf D}^T\cdot{\bf C_{\rm D}}^{-1}\cdot\Delta {\bf D},
\end{equation}
where ${\Delta D}={\bf D}_{\rm obs}-{\bf D}_{\rm th}$, ${\bf D}_{\rm obs}$ and ${\bf D}_{\rm th}$ are the observational and theoretical quantities shown in Table . \ref{tab:bao} and $\bf C_{\rm D}$ is corresponding covariance matrix.

\subsection{CMB data} 
For the CMB dataset, we exploit the parameter of acoustic scale $l_A$, the shift parameter $R$, the decoupling redshift $z_{\ast}$ inferred from the Planck CMB measurement and the distance priors from the Planck 2018  results\citep{collaboration2018planck,2020A&A...641A...5P}, they  are defined respectively as
\begin{equation}
R=\sqrt{\Omega_{\rm m}H^2_0}r(z_\ast),
\end{equation}
\begin{equation}
\ell_{\rm A}=\pi r(z_\ast)/r_{\rm s}(z_\ast),
\end{equation}
where  $\Omega_{\rm{m}}$ is current energy density of dark matter.  $r(z_\ast)$  is the comoving size of the sound horizon at the redshift of the decoupling epoch of photons $z_\ast$, $z_\ast$  can be expressed as \citep{1996ApJ...471..542H,Cao_2011,Lazkoz_2007}:

\begin{equation}
z_\ast=1048[1+0.00124(\Omega_{\rm b}h^2)^{-0.738}][1+g_1(\Omega_{\rm{m}}h^2)^{g_2}],
\end{equation}
where
\begin{equation}
g_1=\frac{0.0783(\Omega_{\rm{b}}h^2)^{-0.238}}{1+39.5(\Omega_{\rm{b}}h^2)^{0.763}},   g_2=\frac{0.560}{1+21.1(\Omega_{\rm{b}}h^2)^{1.81}}.
\end{equation}
The $\chi^2$ distribution for the CMB data can be estimated as
\begin{equation}
\chi^2_{\rm CMB}=\Delta {\bf X}^TC^{-1}_{\rm CMB}\Delta {\bf X},
\end{equation}
where $\omega_b=\Omega_bh^2$ is the baryon energy density, $ h$ is the dimensionless Hubble constant. $\Delta X =\bf X_{obs}- \bf X_{th}$, $\bf X =(R,\ell_{\rm A},\omega_{\rm b})$ and $C_{\rm CMB}$ is the covariance matrix. 

\subsection{H(z) Data}
The $H(z)$ data contain 30 data points in the redshift range from 0 to 2 in Table 4 of \citep{Moresco_2016}, which are obtained using the differential-age method, that is to compare the ages of passively-evolving galaxies with similar metallicity, separated in a small redshift interval \citep{2002ApJ...573...37J}  serving as cosmic chronometers and yielding a model-independent measurement of the expansion rate of the Universe at various redshifts.
The $\chi^2$  distribution for the $H(z)$ data can be expressed as \citep{Cao_2011,Lazkoz_2007}
\begin{equation}
\chi^2_H = \sum_{i=1}^{N=30} \frac{\left[ H_{\rm obs}(z_i)-H_{\rm th}(z_i) \right]^2}{\sigma_H^2}.
\end{equation}
Here $H_{\rm obs}$ and $H_{\rm th}$ are the observational and theoretical Hubble parameters, respectively, and $\sigma_H$ is the error.

Finally, the joint $\chi^2$ of above four datasets employed in this work  can be calculated by
\begin{equation}
\chi^2 = \chi^2_{ SN} + \chi^2_{BAO} + \chi^2_{CMB}+\chi^2_{H(z)}.
\end{equation}

\begin{table}
	\begin{center}
		\caption{The table lists the priors on the parameters space.\label{tab:prior}}
		\begin{tabular}{c|c}
			\hline 
			Parameter&Prior range\\
			\hline
			
			$\Omega_m$&[0, 0.7] \\
			$h$&[0.5, 1]\\
			$\alpha$&[-1, 1]\\
			$\beta$&[-1, 1]\\
			$\ w_d^{eff}$&[-3,-0.3] \\
			$\sigma_8$&[0.2,1.4]\\
			\hline
		\end{tabular}
	\end{center}
\end{table}	

\section{Results}\label{res}
In order to analyse the IDE1 and IDE2 model introduced in Sec.\ref{mod} and place constraints on the free parameters within the scenarios, we use CMB data and the combination of CMB+Pantheon, CMB+BAO+H(z), and CMB+Pantheon++BAO+H(z) datasets and modify the Boltzmann code CLASS \citep{2011JCAP...07..034B}\footnote{\url{https://github.com/lesgourg/class_public}} to solve the background and perturbation equations of  IDE1 and IDE2 model and then we utilize MontePython, a Markov chain Monte Carlo package (MCMC)\citep{Audren:2012wb,Brinckmann:2018cvX,katzgraber2011introduction}\footnote{\url{https://github.com/brinckmann/montepython_public}} with the Metropolis-Hastings algorithm in accordance with the Gelman-Rubin convergence criterion, requiring $\left|R - 1\right| < $ 0.01 to extract the chains of each parameter by computing the reduced chi-square $\chi^2_{red} = \chi^2_{min}/(N-n)$  in the fitting process for each dataset combination, where $\chi^2_{min}$ is the minimum $\chi^2$, $N$ and $n$ are the number of data point and free parameter,respectively. The priors on the parameter space are shown in Table.\ref{tab:prior}.
\begin{figure*}
	\begin{center}
		\includegraphics[width=0.8\textwidth]{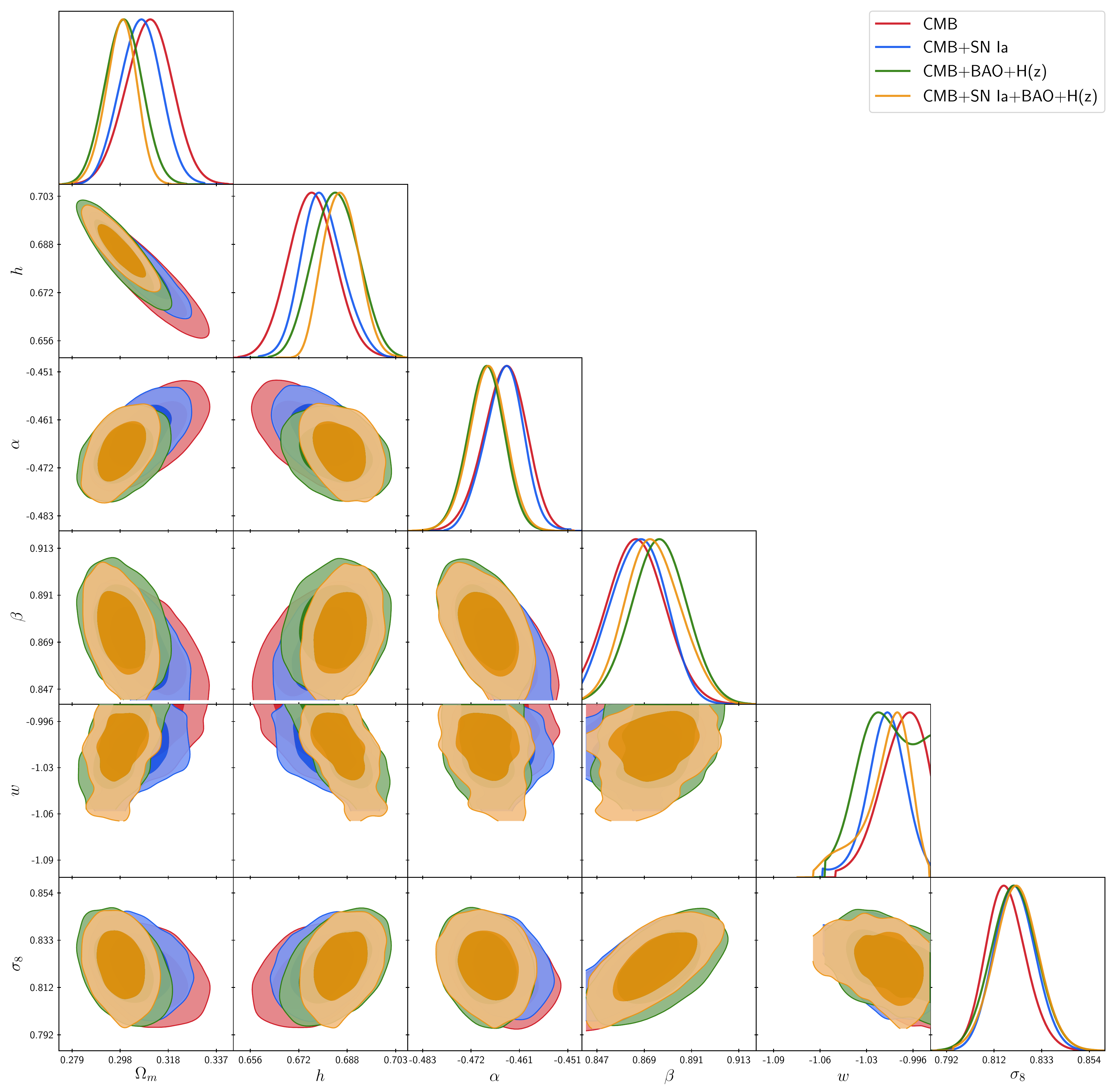}
		\caption{\label{figs:IDE1} The one-dimensional posterior distributions and two-dimensional contours in 1-$\sigma$ (68.3\%) and 2-$\sigma$ (95.5\%) confidence level(C.L.) of the free parameters for the $Q_1 = 3H\alpha\rho_d(1+\beta ln(1-\frac{\rho_c}{\rho_d})$ interacting dark energy scenario. The red, blue, green, and orange plots stand for the constraint results from CMB, CMB+SN Ia, CMB+BAO+H(z) and joint datasets, respectively.}
	\end{center}
\end{figure*}

\begin{table*}
	\begin{center}
		\caption{The table summarizes the mean values and 1-$\sigma$(68.3\%) uncertainties of the parameters under the scenario of the IDE1:$Q_1 = 3H\alpha\rho_d(1+\beta ln(1-\frac{\rho_c}{\rho_d})$ case in Fig.\ref{figs:IDE1}. Here the parameter $\Omega_m$ equal the energy fractions of baryons plus dark matter, i.e. $\Omega_m = \Omega_c +\Omega_b$.}\label{tab:IDE1}
		\begin{tabular}{c|c|c|c|c}
			\hline\hline 
			Parameter&CMB& CMB+SN Ia &CMB+BAO+H(z)&CMB+SN Ia+BAO+H(z) \\
			\hline
			
			$\Omega_m$&$0.322_{-0.009}^{+0.01}$&$0.303_{-0.007}^{+0.007}$&$0.297_{-0.008}^{+0.008}$&$0.298_{-0.006}^{+0.006}$\\
			$h$&$0.677_{-0.008}^{+0.007}$&$0.726_{-0.0067}^{+0.0061}$&$0.691_{-0.007}^{+0.007}$&$0.684_{-0.006}^{+0.0051}$\\
			$\alpha$&$-0.457_{-0.005}^{+0.005}$&$-0.465_{-0.004}^{+0.004}$&$-0.47_{-0.004}^{+0.004}$&$-0.468_{-0.004}^{+0.003}$\\
			$\beta$&$0.861_{-0.014}^{+0.012}$&$0.886_{-0.012}^{+0.013}$&$0.876_{-0.013}^{+0.013}$&$0.874_{-0.013}^{+0.013}$\\
			$\ w_d^{eff}$&$-0.986_{-0.189}^{+0.026}$&$-1.005_{-0.019}^{+0.019}$&$-1.006_{-0.024}^{+0.027}$&$-1.015_{-0.022}^{+0.020}$\\
			$\sigma_8$&$0.811_{-0.009}^{+0.01}$&$0.814_{-0.009}^{+0.008}$&$0.827_{-0.01}^{+0.01}$&$0.822_{-0.01}^{+0.01}$\\
			$\chi^2_{red}$&0.989&1.067&1.173&1.114\\
			\hline
		\end{tabular}
	\end{center}
\end{table*} 

\begin{figure*}
	\begin{center}
		\includegraphics[width=0.8\textwidth]{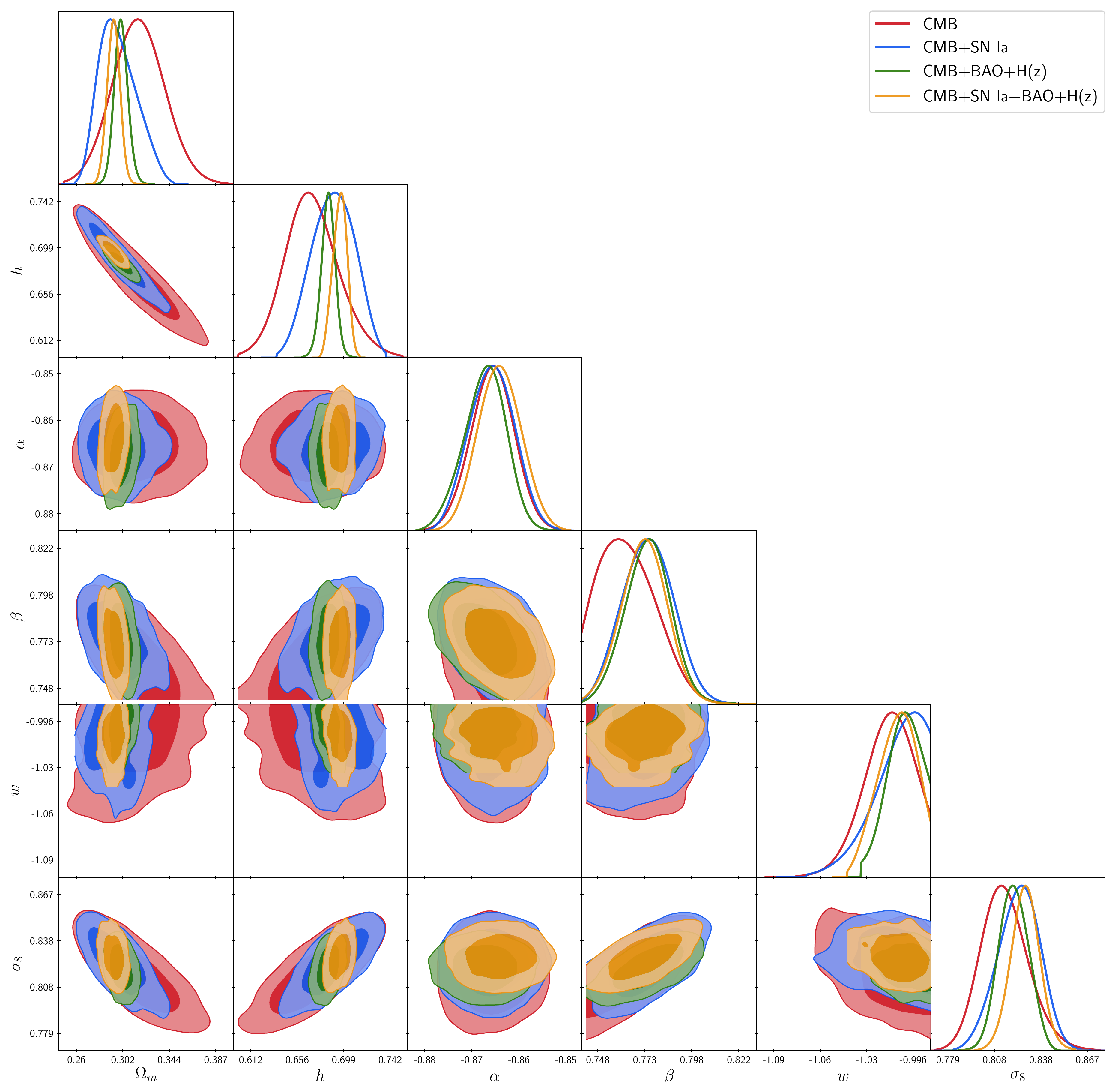}
		\caption{The one-dimensional posterior distributions and two-dimensional contours in 1- $\sigma$ (68.3\%) and 2-$\sigma$ (95.5\%) confidence level(C.L.) of the free parameters for the $Q_2= 3H\alpha\rho_d(1+\beta sin(\frac{\rho_c}{\rho_d}))\rho_c$ interacting dark energy scenario. The red, blue, green, and orange plots show the constraint results from CMB, CMB+SN Ia, CMB+BAO+H(z) and joint datasets, respectively.}\label{figs:IDE2}
	\end{center}
\end{figure*}

\begin{table*}
	\begin{center}
		\caption{The table shows mean values and 1-$\sigma$(68.3/\%) confidence level(C.L.) of the cosmological parameters under the IDE2:$Q_2= 3H\alpha\rho_d(1+\beta sin(\frac{\rho_c}{\rho_d}))$ case in Fig.\ref{figs:IDE2} fitted by above datasets,here $\Omega_m = \Omega_c +\Omega_b$.}\label{tab:IDE2}
		\begin{tabular}{c|c|c|c|c|c}
			\hline\hline 
			Parameter&CMB& CMB+SN Ia &CMB+BAO+H(z)&CMB+SN Ia+BAO+H(z)& \\
			\hline
			$\Omega_m$&$0.317_{-0.022}^{+0.025}$&$0.298_{-0.018}^{+0.015}$&$0.300_{-0.007}^{+0.006}$&$0.294_{-0.006}^{+0.006}$\\
			$h$&$0.680_{-0.028}^{+0.022}$&$0.701_{-0.020}^{+0.020}$&$0.694_{-0.006}^{+0.006}$&$0.686_{-0.006}^{+0.006}$\\
			$\alpha$&$-0.866_{-0.0049}^{+0.0049}$&$-0.866_{-0.005}^{+0.005}$&$-0.867_{-0.0045}^{+0.0047}$&$-0.864_{-0.005}^{+0.005}$\\
			$\beta$&$0.765_{-0.019}^{+0.011}$&$0.775_{-0.013}^{+0.014}$&$0.775_{-0.012}^{+0.013}$&$0.7727_{-0.013}^{+0.012}$\\
			$w_d^{eff}$&$-1.005_{-0.017}^{+0.019}$&$-0.995_{-0.027}^{+0.024}$&$-0.998_{-0.013}^{+0.018}$&$-1.007_{-0.013}^{+0.016}$\\
			$\sigma_8$&$0.815_{-0.017}^{+0.013}$&$0.8236_{-0.013}^{+0.014}$&$0.820_{-0.009}^{+0.009}$&$0.8278_{-0.009}^{+0.009}$\\
			$\chi^2_{min}$&0.973&1.082&1.154&1.203\\
			\hline
		\end{tabular}
	\end{center}
\end{table*} 
Firstly, Fig.\ref{figs:IDE1} and Fig.\ref{figs:IDE2}  show the 68.3\% and 95.5\% confidence level(C.L.) contour map and 1-D PDFs of  the free parameter of IDE1: $Q_1 = 3H\alpha\rho_d(1+\beta ln(1-\frac{\rho_c}{\rho_d})$ and IDE2: $Q_2= 3H\alpha\rho_d(1+\beta sin(\frac{\rho_c}{\rho_d}))$ models
explored by using CMB and CMB+SN Ia and CMB+BAO+H(z) and joint datasets respectively because of these datasets can break the degeneracy of the among quantities. We find that the model can be constrained by these datasets well due to $\chi^2_{red}$ is close to 1 for the above four data combinations. As listed on table \ref{tab:IDE1}, the best-fit value of the coupling factor $\alpha$ and $\beta$ in Fig \ref{figs:IDE1} yield $\alpha \simeq -0.468_{-0.004}^{+0.003}$ and $\beta \simeq 0.874_{-0.013}^{+0.013}$ for the joint datasets and the $Q_1 \approx - 0.00002H(z)<0 $ manifests a difference from zero, i.e. energy transfers from dark matter into dark energy. In addition, the equation of state $w_d^{eff} < -1$ for DE is preferred from the fitting consequences  of CMB+SN Ia+BAO+H(z) combination. The results of other cosmological parameters from the fitting  are also perfectly consistent with observational values. Otherwise, concerning Fig.\ref{figs:IDE2} and Table.\ref{tab:IDE2} present the results  of IDE2: $Q_2= 3H\alpha\rho_d(1+\beta sin(\frac{\rho_c}{\rho_d}))$ scenario, the mean values of the interaction factors  $\alpha$ $\simeq -0.866$ and $\beta \simeq 0.775$  and so $Q_2 \approx 0.00004H(z) > 0 $  meaning the dark energy decays into dark matter in light of the results from observational dataset combinations.

Furthermore, we analyse either issues of  $H_0$ tension or the $\sigma_8$ discrepancy \citep{sym10110585,2013MNRAS.430.2200K} mentioned in Section.\ref{intro} and both issues can be slightly alleviate due to the constraints values under the two interacting  IDE1 and IDE2 framework. From Table.\ref{tab:IDE1} we see, the IDE1 model gives the dimensionless Hubble constant $h$ from 0.666 to 0.686 using the four data combinations, which are almost consistent in 1$\sigma$ level and similarly, the resulting values $\sigma_8$ are tightly obtained from 0.811 to 0.827. While in the IDE2 model, the tensions are also decreased to some extent, the values of $h$ and $\sigma_8$ show no serious deviation from the CMB data only to the joint dataset fitting. In short, these results of parameters from solely Planck CMB data to four observations combinations imply that the non-linear interactions of DE and DM considered in this work can be supported  well by the available observational data which determined from $\Lambda$CDM cosmology.
\begin{figure}
	\centering
	\includegraphics[width=0.6\textwidth]{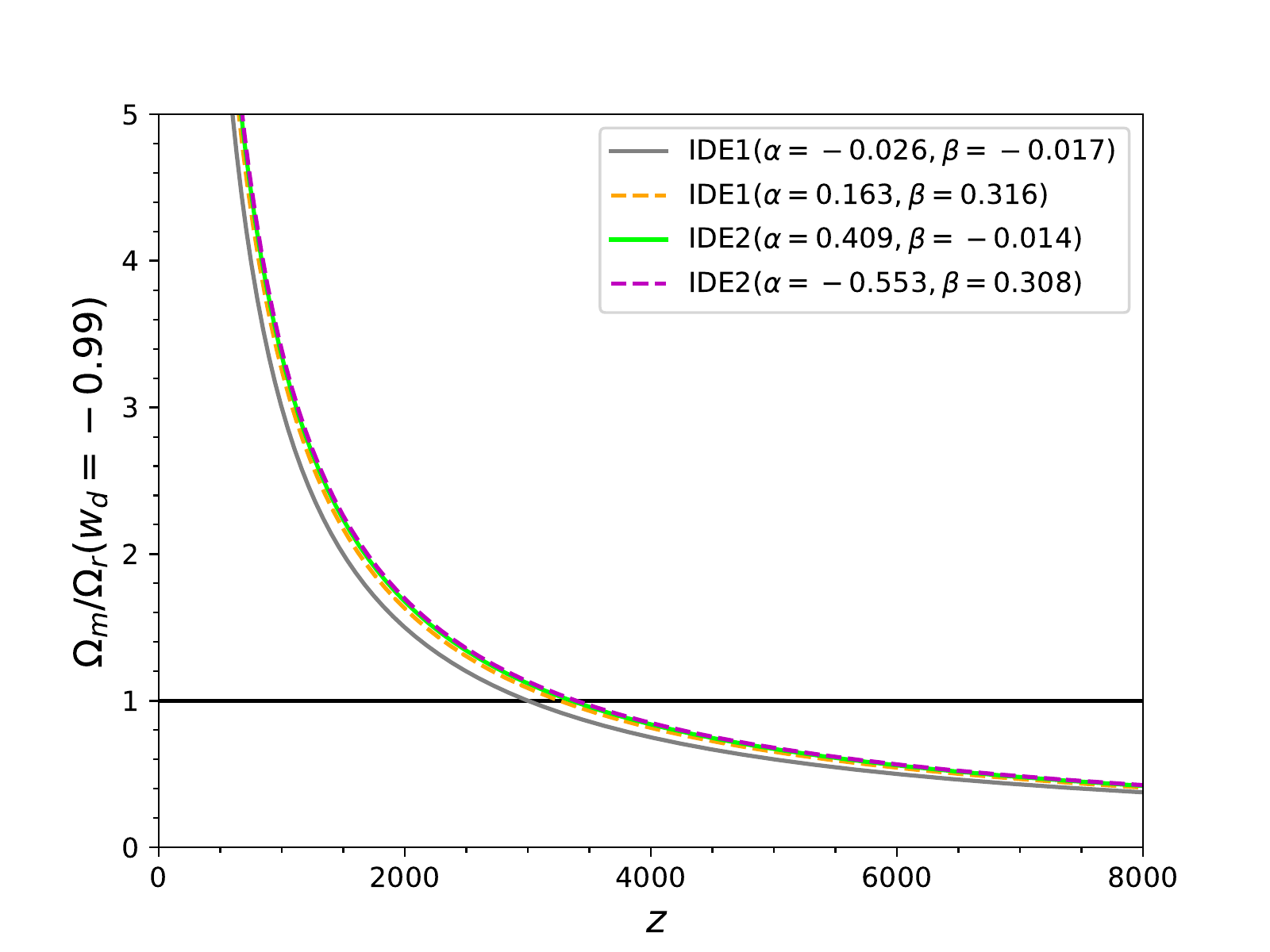}
	\caption{\label{fig:ratio-neuhi}The evolution for the ratio of matter and radiation
		$\Omega_{\rm m}$ and $\Omega_{\rm r}$  with the variation of the interaction parameter $\alpha$ and $\beta$ for IDE1 and IDE2, here $\Omega_{\rm m}= \Omega_{c}+ \Omega_{\rm b}$. The horizontal gray thick line corresponds to the case of
		$\Omega_{\rm m} =\Omega_{\rm r}$, namely the era of the matter-radiation equality
		and we fix these parameters for the plot $\Omega_{m} =0.278$,$\Omega_{b}=0.04$,$\Omega_{d}=0.682$,$\Omega_{\rm r}=0.00005$.}.
	
\end{figure}

In order to further understand the relation between the coupling factor $\alpha$ and $\beta$ and the time of matter-radiation equality for  the existing of an interaction in cosmic dark sector, we draw the evolution curves of $\Omega_{\rm m}/\Omega_{\rm r}$ in redshift space $z$, the time of matter-radiation equality comes later in both IDE1 and IDE2 case with increasing value of $\alpha$ and $\beta$ around redshift $z$=3400 to 3500 compared to non-interacting model as shown in Fig.\ref{fig:ratio-neuhi}, indicating the increasing of interaction parameter $\alpha$ and $\beta$ will lead to a  addition of energy density of the dust matter $\Omega_{\rm m}$, the epoch of matter-radiation balance will become late and the sound horizon will scales up.

\begin{figure*}
	\centering
	\includegraphics[width=0.46\textwidth]{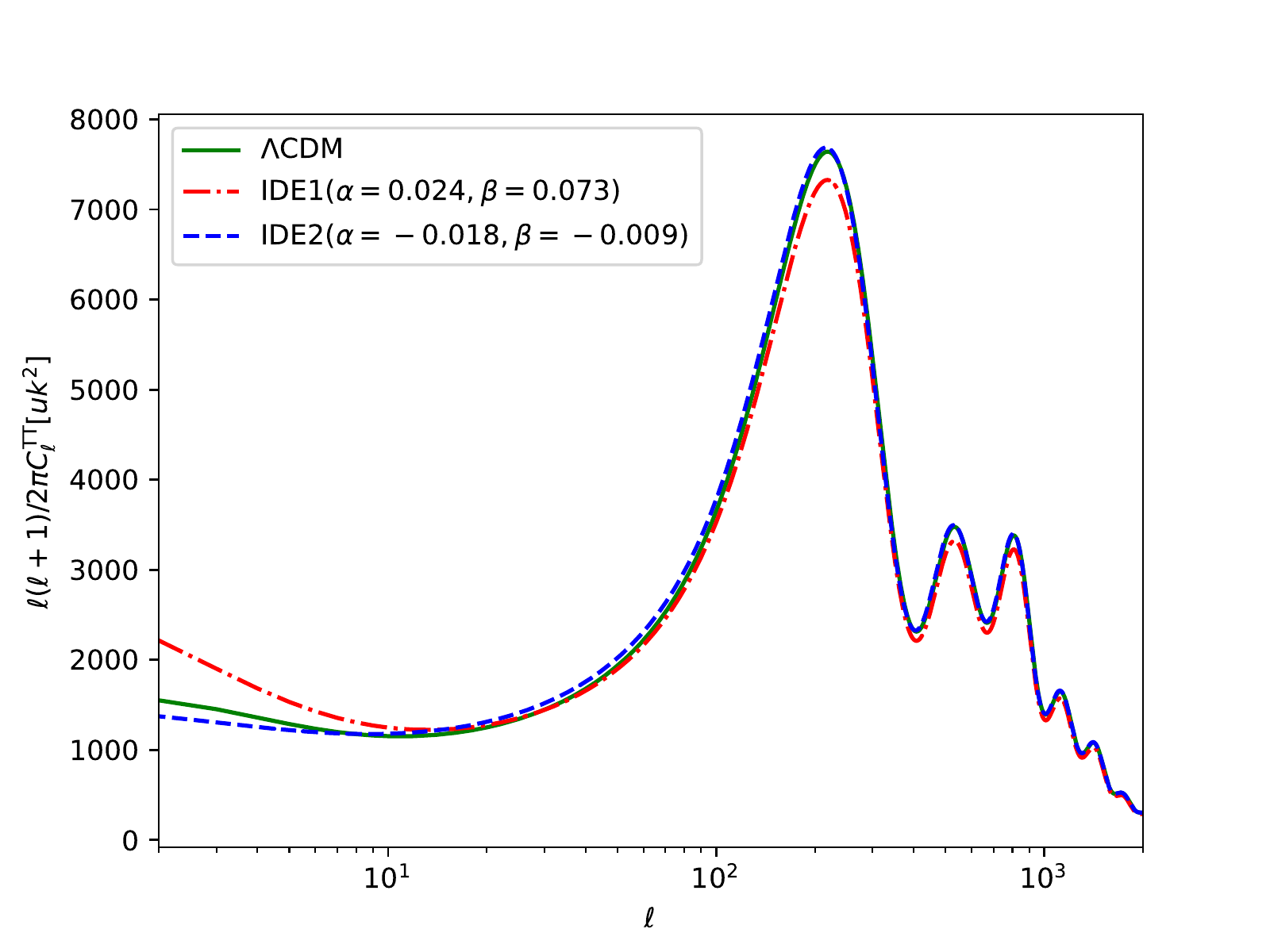}
	\includegraphics[width=0.46\textwidth]{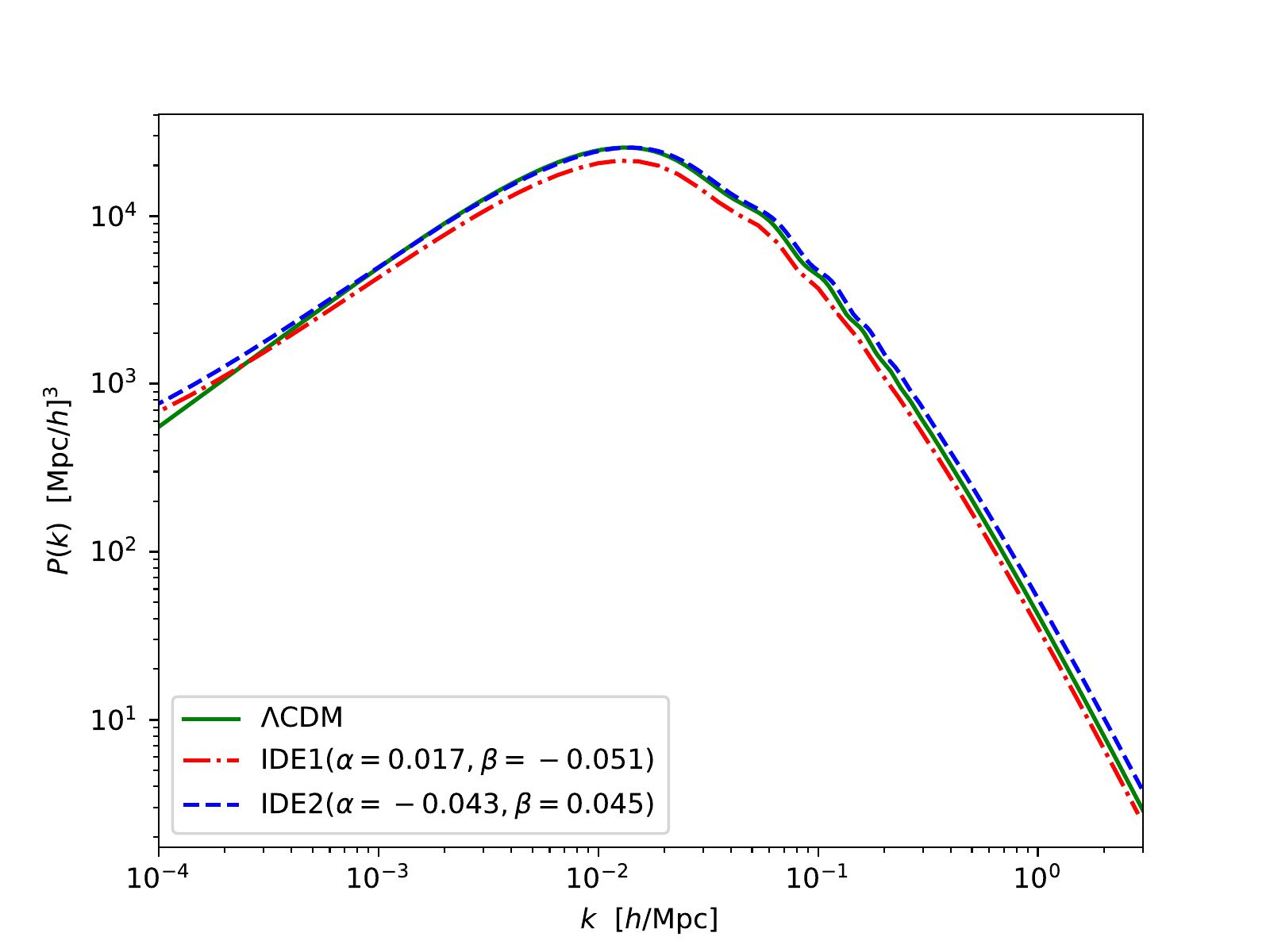}
	\caption{\label{figs:cp}The effects on the CMB temperature power spectra(left panel) and the matter power spectra (right panel)for two case of the interaction term $Q_1$ and $Q_2$. The red and blue curves lines for the IDE1: $Q_1$ and IDE2: $Q_2$ reference to $\Lambda$CDM in green solid line, respectively.}
\end{figure*}
Fig.\ref{figs:cp} presents some key signatures on the CMB temperature power spectrum in left  and the linear matter power spectrum in right panel when taking the different values for $\alpha$ and $\beta$ describing interacting models IDE1 and IDE2. Concerning  the CMB temperature power spectra in the left panel, we deem that actually either dark energy or dark matter gains additional energy density and influences the microwave background temperature anisotropy especially at low multiple $l<30$ under the IDE1 and IDE2 respectively. The increase of dark energy density within IDE1 compared to IDE2, especially alters integrated Sachs-Wolfe(ISW) level on large scale due to the decay of gravitational potential over expansion history as well as changes era of matter-radiation balance and the height of the first peak of CMB temperature power spectra is lifted for the tight coupling of dark sector. As for the right panel, the dominant effects on the linear matter power spectrum $P(k)$ show up at large scales($k<0.042h/Mpc$), hinting the enhancement of the  content of dark matter that forms larger cosmic structures and earlier matter clusters of galaxies under its gravity. Consequently, the shape of matter power spectra $P(k)$ somewhat is more intensified due to larger horizon boundary in IDE2 than IDE1 model. Moreover, the first peak evidently is rised for IDE1 in small multipole $\ell$ <10 since the dark energy fluid flows to dark matter to change the energy of CMB photons when across it. The presence of interaction of  dark sector  changes the regular evolution law and which is imprinted on the integrated Sachs-Wolfe (ISW) effect in large scale effect of the CMB TT spectrum.
\begin{figure*}
	\centering
	\includegraphics[width=0.46\textwidth]{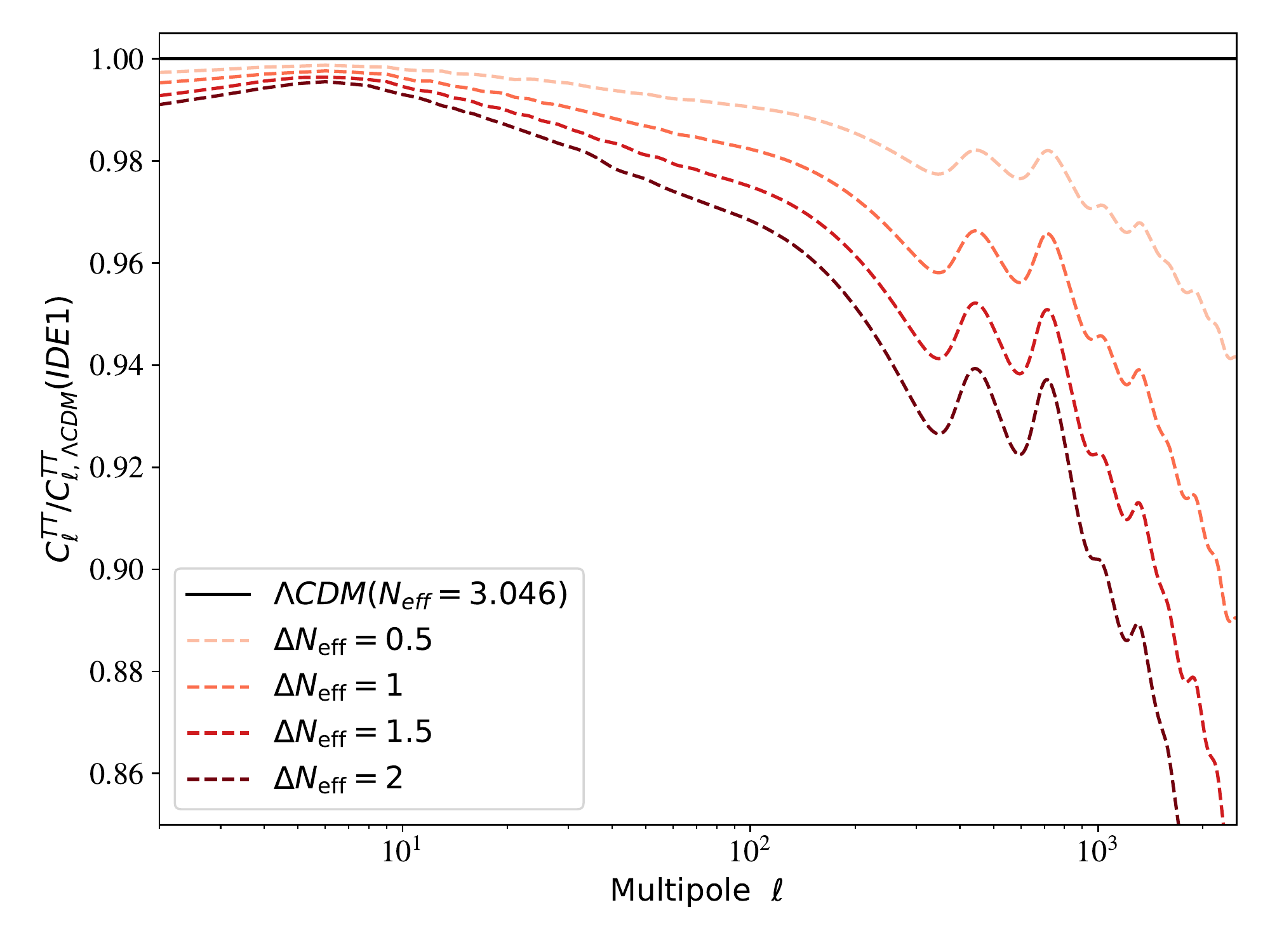}
	\includegraphics[width=0.46\textwidth]{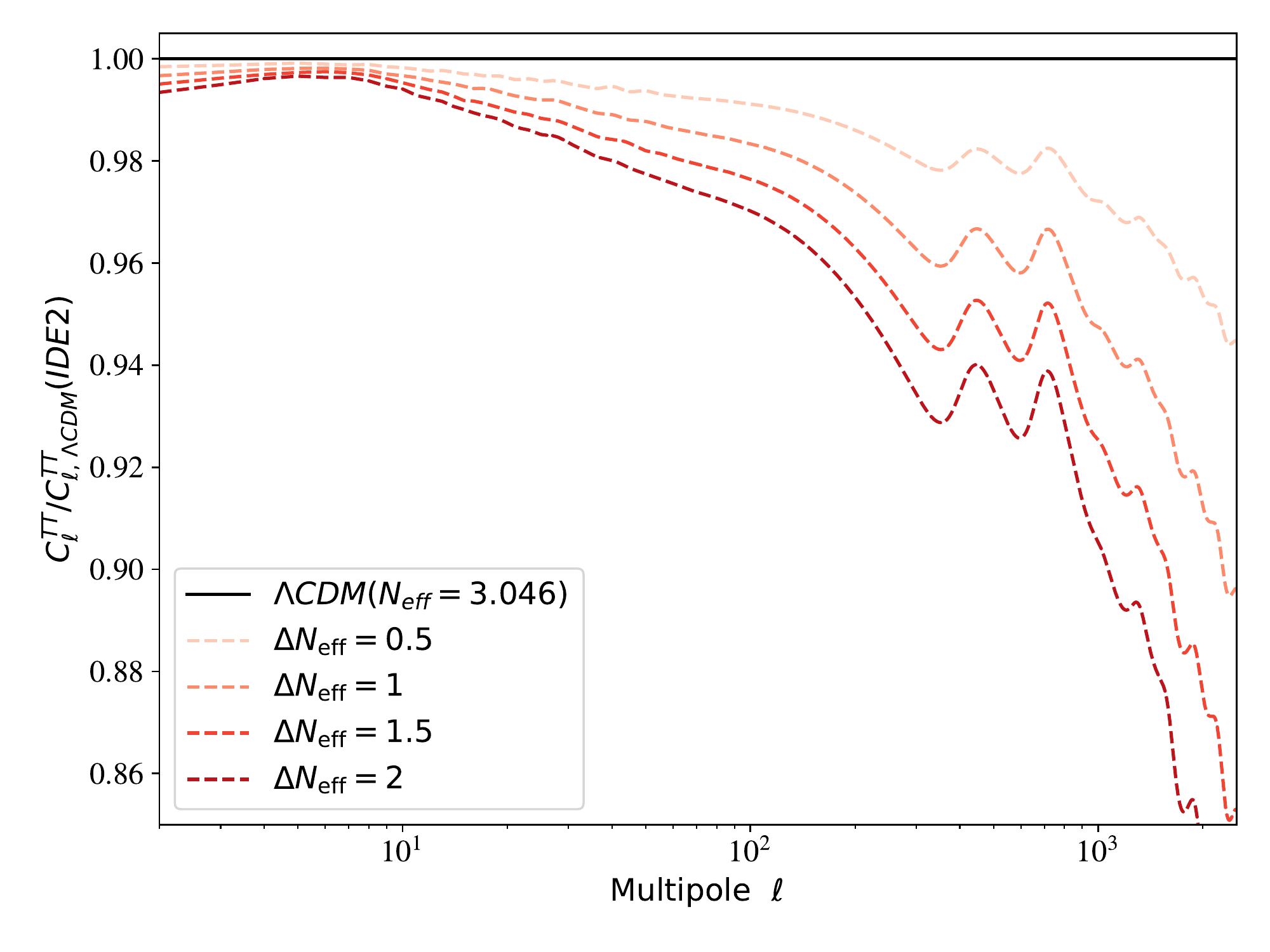}
	\caption{\label{figs:neu-c} The ratio of the CMB temperature power spectra  to $\Lambda$CDM in presence of neutrino for the interaction term IDE1(left panel) and IDE2(right panel) with the variation of $\Delta N_{eff}$ from  0.5 to 2.}
\end{figure*} 
\begin{figure*}
	\centering
	\includegraphics[width=0.46\textwidth]{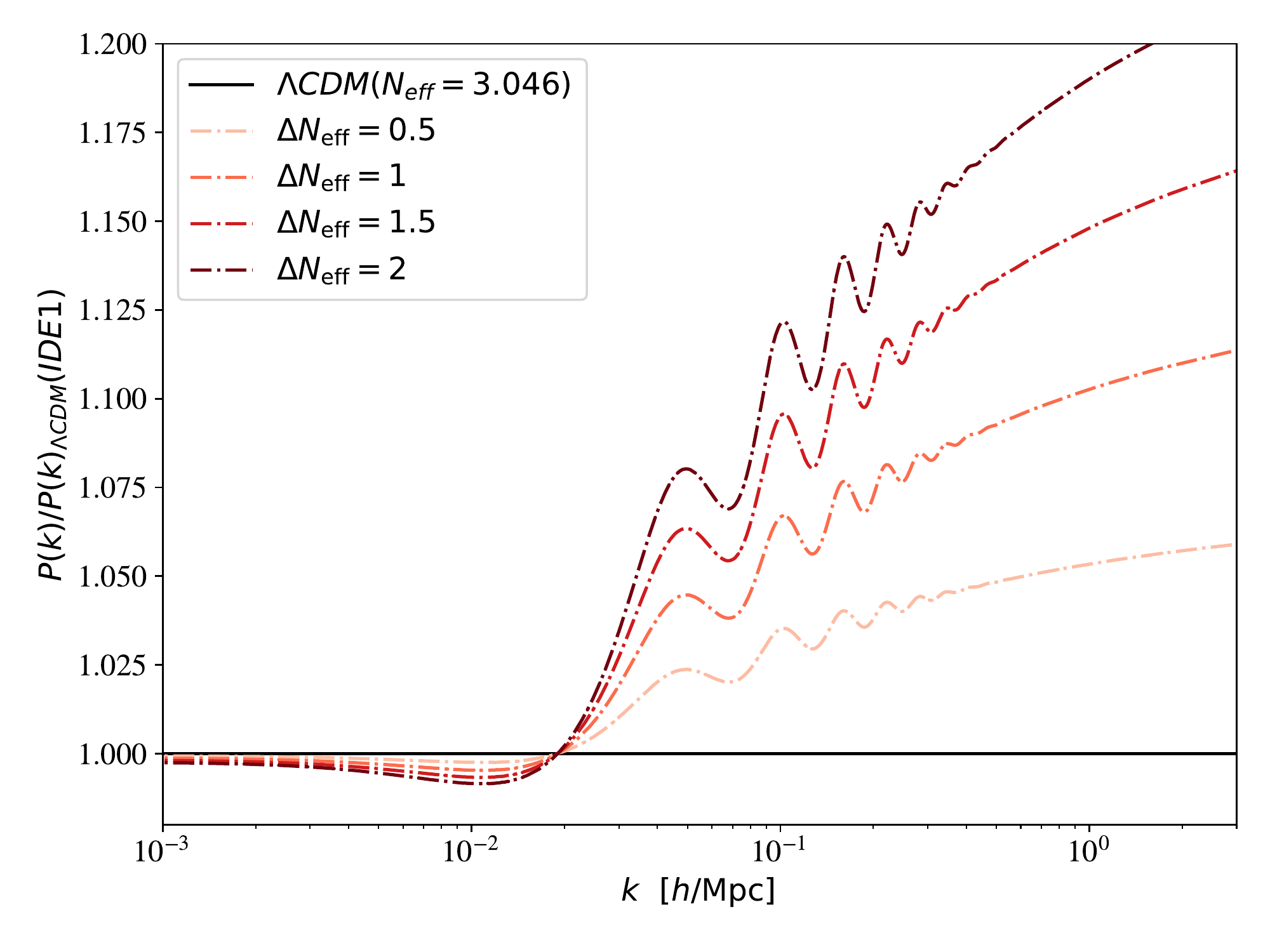}
	\includegraphics[width=0.46\textwidth]{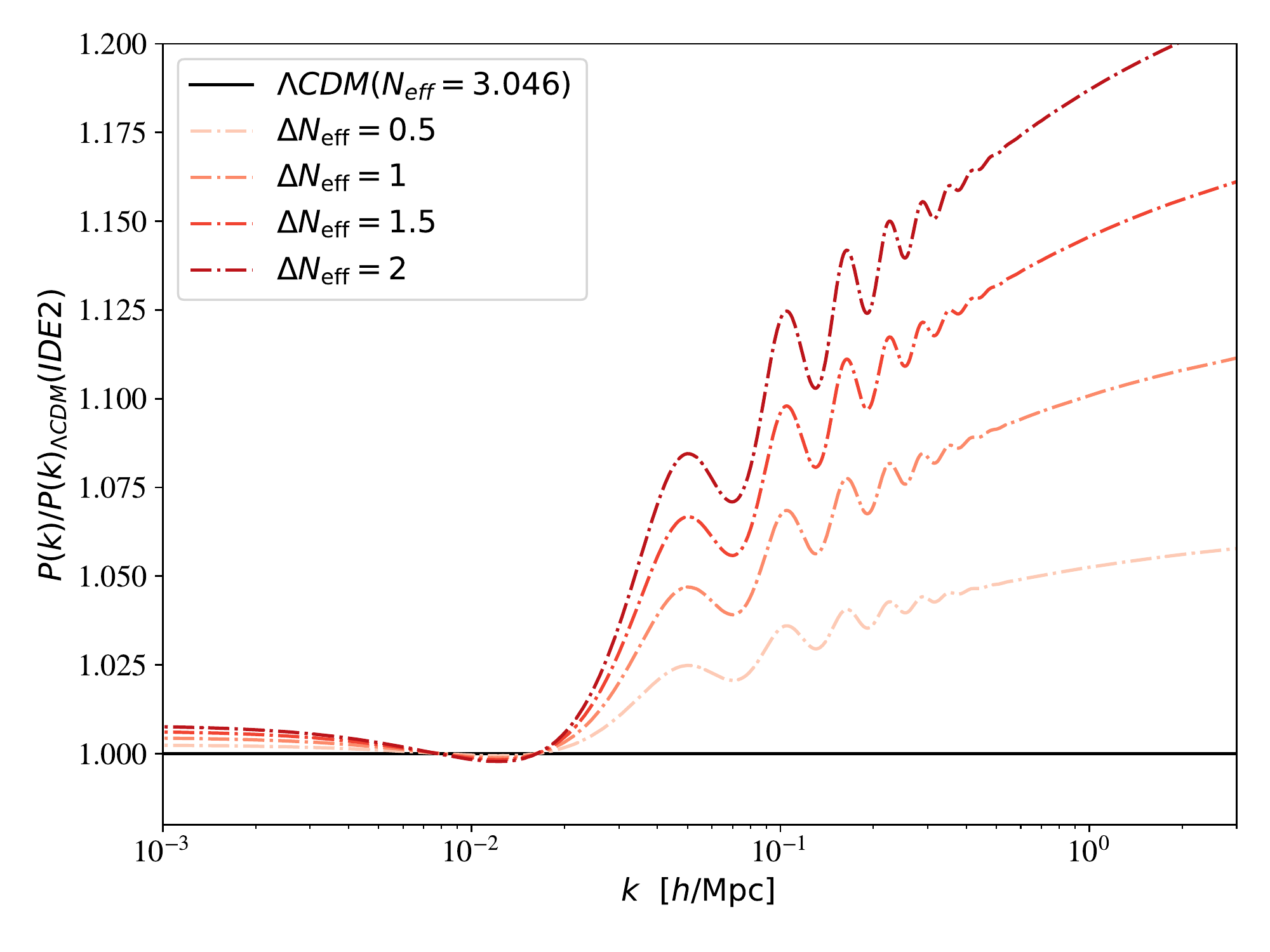}
	\caption{\label{figs:neu-p}The ratio of the matter power spectra to $\Lambda$CDM of the interaction term IDE1(left panel) and IDE2(right panel) when $\Delta N_{eff}$ from 0.5 to 2.}
\end{figure*}

In this paragraph, we comment on neutrino hierarchies problem under the framework IDE1 and IDE2 model. The experiment of neutrino oscillation has suggested that neutrinos have mass size and mass splittings among three flavors neutrino, though either absolute mass of neutrinos or mass hierarchies remian puzzling(see \citep{doi:10.1080/00107510701770539,2018PhRvD..98h3501V,2013arXiv1307.4738B, 2005PhRvD..72l3523B,2020SCPMA..6320401F} for a  review). Actually, both the properties of massive neutrinos and  dark energy could imprint important signatures on expansion history and large scale structure  (LSS) in the evolution history of the universe\citep{2020arXiv201000248Y,2020arXiv200914220C,2001hep.ex....9033K,2020arXiv200914220C,2013arXiv1307.4738B}. Therefore, It's expected that some significant information can be found on $C_{\ell}$ and $P(k)$ spectrum in the direct interaction of dark energy and dark matter scenario when considering mass hierarchies. From  the ratio of $C_{\ell}^{TT}$ to $C_{\ell}^{TT}$ in $\Lambda$CDM model in Fig.\ref{figs:neu-c}, one can find there is mass size and mass hierarchies of neutrino species and showing bigger deviation in IDE1 than IDE2 related to $\Lambda$CDM regardless of NH or IH mode and demonstrating the gradual suppression for  the larger free degree of $\Delta N_{eff}= N-3.046$ from 0.5 to 2 occurring in IDE1 and IDE2 due to the smoothing effect of more neutrinos free screaming. Similarly, concerning the ratio of matter power $P(k)/P(k)_{\Lambda}$CDM  is dramatically intensified in $k>0.02 h/Mpc$ in the wake of $\Delta N_{eff}$ turns bigger from 0.5 to 2 in Fig.\ref{figs:neu-p}, the attribute of neutrinos may strongly degeneracy with dark matter halos to cluster of baryon particles and host a vast of galaxies or other objects in both IDE1 and IDE2 models, but in $k<0.02 h/Mpc$ phase, the magnitude of ratio amplitude  of IDE2 than IDE1 is less than and tend to 1 due to the density of neutrinos are diluted since in larger scales and difficult to cluster.

\section{Conclusions}\label{cons}
 	In this analysis, we have investigated two scenarios of direct interaction between the dark energy and dark matter, in which the interaction term $Q$ is reconstructed as Eq.\ref{eq:IDE1} and Eq.\ref{eq:IDE2} with coupling parameter  $\alpha$ and $\beta$ determining the conversion between DE and DM. 
 
 	Firstly, we come up with  two novel interacting model IDE1 and IDE2, and then
 	we use (i) the cosmic microwave background from Planck 2018 result, (ii) Pantheon sample of Supernovae Type Ia, (iii) baryon acoustic oscillations distance measurements and (iv) direct H(z) observation to  put constraints on the two interacting scenarios $Q_1 = 3H\alpha\rho_d(1+\beta ln(1-\frac{\rho_c}{\rho_d})$ and $Q_2= 3H\alpha\rho_d(1+\beta sin(\frac{\rho_c}{\rho_d}))$ by calling the MontePython sampler and Class code. All results summarized in Table.\ref{tab:IDE1} and \ref{tab:IDE2} and Fig.\ref{figs:IDE1} and \ref{figs:IDE2}, we find that 
 	
 	(1) our analysis with all combined datasets support non-zero value for the interacting factors $\alpha$ and $\beta$ while the dark energy equation of state(EoS) $w< -1$ at  2$\sigma$ confidence level.
 	
 	(2) With IDE1 or IDE2 model, we point out that the results from the interactions are compatible with a general model such as the uncoupled $w$CDM cosmology and fit the latest observations. Specially, it is evident that the interacting models relaxed the existing tensions of $H_0$ and $\sigma_8$ arised from different cosmological measurements relying on the $\Lambda$CDM picture at about 2$\sigma$ confidence level, though it does not seem to be able to completely alleviate them. 
 	
 	(3) When the addition of neutrinos mass normal hierarchy(NH) and inverse hierarchy(IH) mode  to the interaction theory of IDE1 and IDE2 scenario, we discuss that the influence on the ratio of CMB temperature spectrum, indicating the neutrinos obviously smooth temperature fluctuation especially in $k>100$ with the free degree error $\Delta N_{eff}$ from 0.5 to 2, while the effect of dark matter to drag baryons particles to cluster and spark the physical processes are reinforced due to the more neutrinos behaving like  dark matter to form larger structures of halos. 
 	
 	As a whole, we confirm that the interaction  of dark energy and dark matter in this work can be explored by the cosmological observation combinations and eliminate some discrepancies  to some extent that from theory and determination and even can produce some key signals on the $C_{\ell}$ and $P(k)$ spectrum can be detected in a number of cosmology surveys in future.

\section*{Acknowlegment}
 We thank Tongjie Zhang for constructive comments on the manuscript.
  
 \bibliography{DMDE}
 \end{document}